\documentclass{svproc}
\usepackage{url}

\usepackage{amsmath}
\usepackage{amssymb}
\usepackage{graphicx}
\usepackage{csquotes}
\usepackage{multirow}
\usepackage{tabularx,booktabs}
\usepackage{xcolor,colortbl} 
\usepackage[colorlinks]{hyperref} 
\usepackage{mathtools}
\usepackage{wasysym}
\usepackage{subcaption}
\usepackage{enumitem}
\usepackage{epstopdf}
\usepackage{epsfig}
\usepackage{rotating}
\usepackage{caption} 
\usepackage{academicons}
\definecolor{orcidlogocol}{HTML}{A6CE39}

\usepackage{amsmath}
\DeclareMathOperator*{\argmax}{arg\,max}
\DeclareMathOperator*{\argmin}{arg\,min}
\definecolor{green}{HTML}{48bf91}
\definecolor{gray}{HTML}{605e5e}
\definecolor{tab_gray}{gray}{0.85}
\newcolumntype{a}{>{\columncolor{tab_gray}}c}

\newcommand{\oneS}{\ensuremath{{}^{\textstyle *}}}
\newcommand{\twoS}{\ensuremath{{}^{\textstyle **}}}

\begin{document}
\mainmatter              
\title{Building resilient organizations: The roles of top-down vs. bottom-up organizing}
\titlerunning{Building resilient organizations}  
%
\author{Stephan Leitner \textsuperscript{[0000-0001-6790-4651]}}
%
\authorrunning{Stephan Leitner} 
%
\tocauthor{Stephan Leitner}
\institute{University of Klagenfurt, Klagenfurt, Austria,\\
\email{stephan.leitner@aau.at}}

\maketitle              

\begin{abstract}
Organizations face numerous challenges posed by unexpected events such as energy price hikes, pandemic disruptions, terrorist attacks, and natural disasters, and the factors that contribute to organizational success in dealing with such disruptions often remain unclear. This paper analyzes the roles of top-down and bottom-up organizational structures in promoting organizational resilience. To do so, an agent-based model of stylized organizations is introduced that features learning, adaptation, different modes of organizing, and environmental disruptions. The results indicate that bottom-up designed organizations tend to have a higher ability to absorb the effects of environmental disruptions, and situations are identified in which either top-down or bottom-up designed organizations have an advantage in recovering from shocks.

\end{abstract}

\section{Introduction}

Organizations face numerous challenges posed by unexpected events, including sudden energy price hikes, pandemic disruptions, terrorist attacks, and natural disasters. Some organizations appear to be more adept than others in handling and recovering from such disruptions, as evidenced by anecdotal accounts \cite{fiksel2015}. However, the factors that contribute to an organization's success in dealing with and recovering from disruptions are not always clear \cite{linnenluecke2017resilience}. 

Resilience, a key concept in organizational research, has gained considerable attention over the last twenty years, leading to varied applications in different sectors and topics \cite{hepfer2022heterogeneity,linnenluecke2017resilience}. The concept traces back to Staw et al. \cite{staw1981threat} and Meyer \cite{meyer1982adapting}, who examined how organizations respond to external threats. After initial focus on internal threats, post-9/11 research shifted to external threats, expanding the application of Staw and colleagues' and Meyer's ideas. Recent studies, especially in supply-chain research, offer varied definitions of resilience. Ponomarov and Holcomb \cite{ponomarov2009understanding} see it as preparation and recovery from unforeseen events, while Christopher and Peck \cite{christopher2004supplychain} view it as bouncing back or improving after a shock, a perspective shared by resilience engineering \cite{hollnagel2011resilienceengineering}.

Research on resilience often operates in isolation across levels (individual, team, organization) and scientific disciplines, hindering the development of integrated research \cite{raetze2021resilience}. As a result, it is challenging to gain a comprehensive understanding of resilience particularly across different levels.

There are, however, some papers that have addressed the issue of multi-level factors in the context of resilience. For instance, Youssef and Luthans \cite{youssef2005} argue that organizational assets may enable organizational resilience, and resilience at the leader level may increase employee resilience. Branicki et al. \cite{branicki2019resilience} propose that individual-level resilience of entrepreneurs might enable organizational resilience. Some studies have provided empirical evidence for multi-level effects of resilience. For instance, Prayag et al. \cite{prayag2020psychological} provide evidence for multi-level resilience in the tourism sector, while McEwen et al. \cite{mcewen2018measure} demonstrate the importance of multi-level resilience for small and medium-sized enterprises.\footnote{For a comprehensive review of the literature on organizational resilience, readers are referred to the work of Raetze et al. \cite{raetze2021resilience} and Linnenluecke \cite{linnenluecke2017resilience}, among others.}

This paper seeks to enhance the current understanding of resilience across different organizational levels. Specifically, this study examines the (macro-level) resilience of, first, newer organizational forms, such as holacracies and self-organized organizations, which share the common feature of relying on different forms of bottom-up task allocation at the micro-level and, second, traditional organizational forms are typically designed top-down, with task allocation being dictated from the top of the organization.
The primary question that this research aims to answer is whether, and if so, under which circumstances, organizations are more resilient to disruptions in the environment. The specific focus of the analysis is on which type of organization can \textit{(i)} better absorb the effects of disruptions in the environment and \textit{(ii)} better recover from such shocks.

This research builds upon previous work on autonomously (bottom-up) designed organizations. In \cite{leitner2023designing}, the author presents the results of an extensive analysis of the interplay between bottom-up task allocation that is governed by short-term or long-term oriented motives, learning at the individual level, and incentives. This work is further extended in \cite{leitner2023collaborative}, where the interplay between different incentive mechanisms and search modes in organizational decision-making is analyzed. Both papers consider disruption-free environments.
The research presented in this paper extends these models by \textit{(i)} incorporating a model of environmental disruptions with external control over the severity of the disruption, and \textit{(ii)} including a more detailed model of the motives that individual agents follow during autonomous task allocation. The findings suggest that top-down designed organizations, compared to their bottom-up designed counterparts, typically experience a more significant decline in organizational performance when disruptions occur. Furthermore, while top-down designed organizations may have an advantage in recovering from disruptions in certain contexts, the more flexible and decentralized structure of bottom-up designed organizations may better equip them to recover from disruptions in a wider range of scenarios.

The remainder of this paper is structured as follows. In Sec. \ref{sec:model}, an agent-based model of organizations with top-down or bottom-up task allocation is introduced. Section \ref{sec:results} is dedicated to presenting and discussing the results on the ability of organizations to absorb and recover from disruptions in the environment. Finally, Sec. \ref{sec:conclusion} summarizes the key findings of this paper

\section{Model}
\label{sec:model}

The stylized model of an organization is based on the $N\!K$ framework \cite{wall2021agent,leitner2015simulation}. Organizations consist of $M \in \mathbb{N}$ departments, each represented by an agent. These agents work together on a performance landscape that represents an $N$-dimen\-sional binary decision problem, with $N \in \mathbb{N}$. The decision-problem's $N$ dimensions are interconnected, with $K \in \mathbb{N}_0$ interdependencies affecting the ruggedness of the landscape on which the agents operate. 

\subsection{Task environment}
\label{sec:landscape}

\subsubsection{Landscapes} The $N$-dimensional binary decision problem can be denoted as $\mathbf{d} = [d_1, d_2, \dots, d_N]$, where $d_n \in \{0, 1\}$ and $n=1,\dots,N$. Each decision $d_n$ contributes $f(d_n) \sim U(0,1) \subset \mathbb{R}$ to the organizational performance. However, due to potential interdependencies, the performance contribution $f(d_n)$ is not solely determined by decision $d_n$ but also by $K \in \mathbb{N}_0$ other decisions. The corresponding pay-off function can be formalized as $f(d_n) = f(d_n, d_{i_1}, \dots, d_{i_K})$, where $\{i_1, \dots, i_K\} \subseteq \{1, \dots, i-1, i+1, \dots, N\}$. The performance $P(\mathbf{d})$ of solution $\mathbf{d}$ is the average of all $N$ performance contributions:
\begin{equation}
\label{eq:performance}
    P(\mathbf{d}) = \frac{1}{|\mathbf{d}|} \sum_{i=1}^{N} f(d_n)~.
\end{equation}
\subsubsection{Disruptions in the environment} While agents operate on the performance landscape, there may be disruptions that affect the performance contributions $f(d_n)$, causing the shape of the landscape to change. This paper specifically considers such shocks, and the model allows for controlling their severity through the correlation parameter $\rho \in (-1, 1) \subset \mathbb{R}$. The correlated performance contributions $f^c(d_n)$ are created following the procedure proposed in \cite{demirtas2014generating}. First, $v_n \sim U(0,1) \subset \mathbb{R}$ and $w_n \sim B(a,1) \subset \mathbb{R}$ are drawn. The shape parameter $a$ of the latter Beta distribution is a function of the correlation parameter:
\begin{equation}
\label{eq:correlation}
a = \frac{1}{2} \left( \sqrt{\frac{49+\rho}{1+\rho}} -5 \right)~.
\end{equation}
Next, $v_n$ and $w_n$ are used to compute the correlated performance contribution $f^c(d_n)$ as follows:
\begin{equation}
\label{eq:correlated-shock}
f^c(d_n) =
\begin{cases}
|w_n - f(d_n)| & \text{if } v_n < 0.5 \\
1 - |1 - w_n - f(d_n)| & \text{if } v_n \geq 0.5~.
\end{cases}
\end{equation}

\subsection{Search process}

The tasks that are assigned to agent $m$ and are represented by $\mathbf{d}_m$, which is referred to as agent $m$'s area of responsibility. The remaining tasks, i.e., those \textit{not} assigned to agent $m$, are referred to as residual tasks and are denoted by $\mathbf{d}_{-m}$. The organization has introduced a linear incentive mechanism that takes into account both the agents' individual and residual performances and uses the incentive parameter $\lambda \in (0, 1) \subset \mathbb{R}$ to weigh them. The utility of agent $m$ in period $t$ can be expressed as:
\begin{equation}
\label{eq:utility}
 U(\mathbf{d}_{mt}, \mathbf{d}_{-mt}) = \lambda \cdot P(\mathbf{d}_{mt}) + (1 - \lambda) \cdot P(\mathbf{d}_{-mt})~.
\end{equation}
%
In each period, the agents aim to improve the performance of their tasks and increase their utility. To achieve this goal, they search for a better solution $\mathbf{d}^{\ast}_{mt}$ to their partial decision problem within the neighborhood of the current solution. The neighborhood is defined by a Hamming distance of 1. During this search process, agents do not communicate with each other. Therefore, they rely on the residual tasks of the previous period, represented by $\mathbf{d}_{-mt-1}$. Based on this, agents make their decision in period $t$ by following the below rule:
\begin{equation}
\mathbf{d}_{mt} = 
\argmax_{ 
    \mathbf{d}^\prime \in \{\mathbf{d}^{\ast}_{mt}, \mathbf{d}_{mt-1}\}} 
    U (\mathbf{d}^\prime, \mathbf{d}_{-mt-1})
 \end{equation}
\noindent At the organizational level, the overall solution is achieved by combining the solutions to the partial problems, $\mathbf{d}_t = \cup_{m=1}^{M} \mathbf{d}_{mt}$, and the corresponding performance can be computed using Eq. \ref{eq:performance}.

\subsection{Learning mechanism}

The agents know that their tasks are interdependent but lack precise knowledge about the value of $K$ and the patterns of these interdependencies. Nevertheless, they utilize their observations to form expectations about these interdependencies. Once the solution $\mathbf{d}_t$ is implemented, all agents can observe its performance contribution in their respective areas of responsibility, denoted by $f(d_{it})$ where $d_{it} \in \mathbf{d}_{mt}$. Agent $m$ maintains a record of the number of instances where they observed interdependence and non-interdependence between task $d_i$ and performance contribution $f(d_j)$ up to period $t$, represented by $\alpha_{mt}^{ij}\in\mathbb{N}$ and $\beta_{mt}^{ij}\in\mathbb{N}$, respectively, where $i,j \in \mathbb{N}$ and $i \neq j$.\footnote{Note that in the first period, $\alpha$ and $\beta$ are set to one, resulting in initial beliefs of $0.5$ as shown in Eq. \ref{eq:belief}.} Over time, the values of $\alpha$ and $\beta$ are updated using the following rule:
\begin{equation}
    (\alpha_{mt}^{ij}, \beta_{mt}^{ij}) = 
    \begin{cases}
        (\alpha_{mt-1}^{ij}, \beta_{mt-1}^{ij}) & \text{if } \mathbf{d}_t = \mathbf{d}_{t-1}~,\\ 
        (\alpha_{mt-1}^{ij}+1, \beta_{mt-1}^{ij}) & \text{if } \mathbf{d}_t \neq \mathbf{d}_{t-1} \text{ and } f(d_{jt}) \neq f(d_{jt-1})~,\\
        (\alpha_{mt-1}^{ij}, \beta_{mt-1}^{ij} + 1) & \text{if } \mathbf{d}_t \neq \mathbf{d}_{t-1} \text{ and } f(d_{jt}) = f(d_{jt-1})~.
    \end{cases}
\end{equation}
Let us represent agent $m$'s expectation regarding the interdependence between decisions $d_i$ and $d_j$ in period $t$ by $\eta_{mt}^{ij} \in \left(0,1\right) \subset \mathbb{Q}$. Specifically, these expectations are defined as the mean of a Beta distribution,
\begin{equation}
    \label{eq:belief}
    \eta_{mt}^{ij} = E(X) = \frac{\alpha_{mt}^{ij}}{\alpha_{mt}^{ij} + \beta_{mt}^{ij}}~, \text{where } X \sim B(\alpha_{mt}^{ij},\beta_{mt}^{ij})~.
\end{equation}

\subsection{Task allocation}
\label{sec:task-allocation}

\subsubsection{Bottom-up task allocation} If organizations are designed bottom-up, the tasks assigned to agent $m$, which are represented by $\mathbf{d}_{mt}$, can change over time as the agents are allowed to reallocate tasks periodically every $\tau$ periods. However, agents are restricted in the number of tasks they can handle at a time, such that the number of tasks assigned to an agent is limited to the interval $(1,C) \subset \mathbb{N}$.  The reallocation process involves several steps. Firstly, agents offer a task that falls within their responsibility areas to other agents. Secondly, all agents are notified of the offers, and they can signal their willingness to take charge of the offered tasks. 

\paragraph{Offering tasks} During the reallocation process, agents take into account two crucial factors when deciding which task to offer: 

\begin{enumerate}[label=(\roman*)]
\item To assess the impact of task reallocation on utility in period $t$, agent $m$ calculates the difference between their utility with and without being responsible for a specific task ${d}_{it} \in \mathbf{d}_{mt}$. The expected change in agent $m$'s utility if task ${d}_{it}$ is assigned to another agent is denoted by
\begin{equation}
\label{eq:delta-utility}
U^{-}{} (\mathbf{d}_{mt}, \mathbf{d}_{-mt}, {d}_{it}) = U({\mathbf{d}_{mt} \setminus \{{d}_{it}\}},\mathbf{d}_{-mt} \cup \{{d}_{it}\}) - U(\mathbf{d}_{mt},\mathbf{d}_{-mt})~,
\end{equation}
where the symbols $\setminus $ and $\cup $ indicate the set difference and union, respectively.
\item The second dimension concerns the extent to which the task allocation aligns with their beliefs regarding interdependencies between tasks. In this perspective, agents aim at maximizing interdependencies within their own areas of responsibility and minimizing the interdependencies of their tasks with the tasks assigned to other agents, a concept known as modularization or \enquote{mirroring hypothesis} \cite{colfer2016mirroring}. To estimate this factor, agent $m$ calculates their belief regarding the interdependencies between task ${d}_{it}$ and the other tasks within their area of responsibility as follows:
\begin{equation}
H(\mathbf{d}_{mt}, {d}_{it}) = \sum_{\substack{\forall j: d_{jt} \in \mathbf{d}_{mt} \\ i \neq j}} \eta^{ij}_{mt}~.
\end{equation}
\end{enumerate}
By considering these two factors, agent $m$ offers the following task to the other agents:
\begin{equation}
\label{eq:offers}
\tilde{d}_{mit} = \argmin_{d^{\prime} \in \mathbf{d}_{mt}} \left( \gamma \cdot \left(1 - U^{-}{} (\mathbf{d}_{mt}, \mathbf{d}_{-mt}, {d}^{\prime})\right) + (1 - \gamma) \cdot H(\mathbf{d}_{mt}, {d}^{\prime})\right) ~,
\end{equation}
where $\gamma \in (0, 1) \subset \mathbb{R}$ reflects the extent to which agents consider these two factors during the reallocation.\footnote{It is worth noting that Eq. \ref{eq:offers} returns the argument that \textit{minimizes} the function. As agents aim to \textit{maximize} their utility and performances are normalized to one, the utility is reflected by $(1 - U^{-}{} (\mathbf{d}_{mt}, \mathbf{d}_{-mt}, {d}^{\prime}))$ in Eq. \ref{eq:offers}.} It's important to note that agents are only able to offer tasks if the number of tasks assigned to them, denoted by $|\mathbf{d}_{mt}|$, is greater than one.

\paragraph{Signals and task reallocation} After all agents have identified the tasks they want to reallocate, they are notified of all available tasks. If an agent, say agent $m$, has free resources, meaning $1 \leq |\mathbf{d}_{mt}| < C$, they express their interest in all tasks by sending a signal. The signal for task $\tilde{d}_{nit}$ that agent $m$ sends, where $n=1\dots,M$, is determined by the following rule:
\begin{equation}
    s_m(\tilde{d}_{nit}) = \gamma \cdot U^{+}{} (\mathbf{d}_{mt}, \mathbf{d}_{-mt}, \tilde{d}_{nit}) + (1 - \gamma) \cdot H(\mathbf{d}_{mt}, \tilde{d}_{nit})~, 
\end{equation}
where 
\begin{equation}
\label{eq:delta-plus}
U^{+}{} (\mathbf{d}_{mt}, \mathbf{d}_{-mt}, \tilde{d}_{nit}) = U({\mathbf{d}_{mt} \cup \{\tilde{d}_{nit}\}},\mathbf{d}_{-mt} \setminus \{\tilde{d}_{nit}\}) - U(\mathbf{d}_{mt},\mathbf{d}_{-mt}) + \varepsilon~.
\end{equation}
To account for agents' imperfect ability to predict the performance contributions of tasks outside their area of responsibility, Eq. \ref{eq:delta-plus} includes an error term denoted as $\varepsilon$, which follows a normal distribution with mean $0$ and variance $0.01$. After all agents have sent their signals, tasks are reassigned to the agent with the highest signal for each task. However, if the highest signal for task $\tilde{d}_{mit}$ is offered by agent $m$ and has a value of $s_m(\tilde{d}_{mit})$, the task is not reassigned, and agent $m$ continues to be responsible for the task until the next round of reallocation.

\subsubsection{Top-down task allocation} In the case of top-down designed organizations, it is assumed that a \enquote{central organization designer} is aware of $K$ and the pattern of interdependencies between tasks, and this designer allocates tasks to agents to minimize cross-interdependencies between departments. This approach is in line with the more classical school of thought of organizational design. In this paper, this can be achieved by a symmetric and sequential task allocation, meaning that every agent is in charge of the same number of tasks, and agent 1 is responsible for the first $N/M$ tasks, agent 2 is in charge of tasks the next $N/M$ tasks, etc. 

\begin{figure*}
    \centering
    \begin{subfigure}[t]{0.48\textwidth}
        \label{fig:matrices-decomposable}
        \centering
        \includegraphics[scale=0.30]{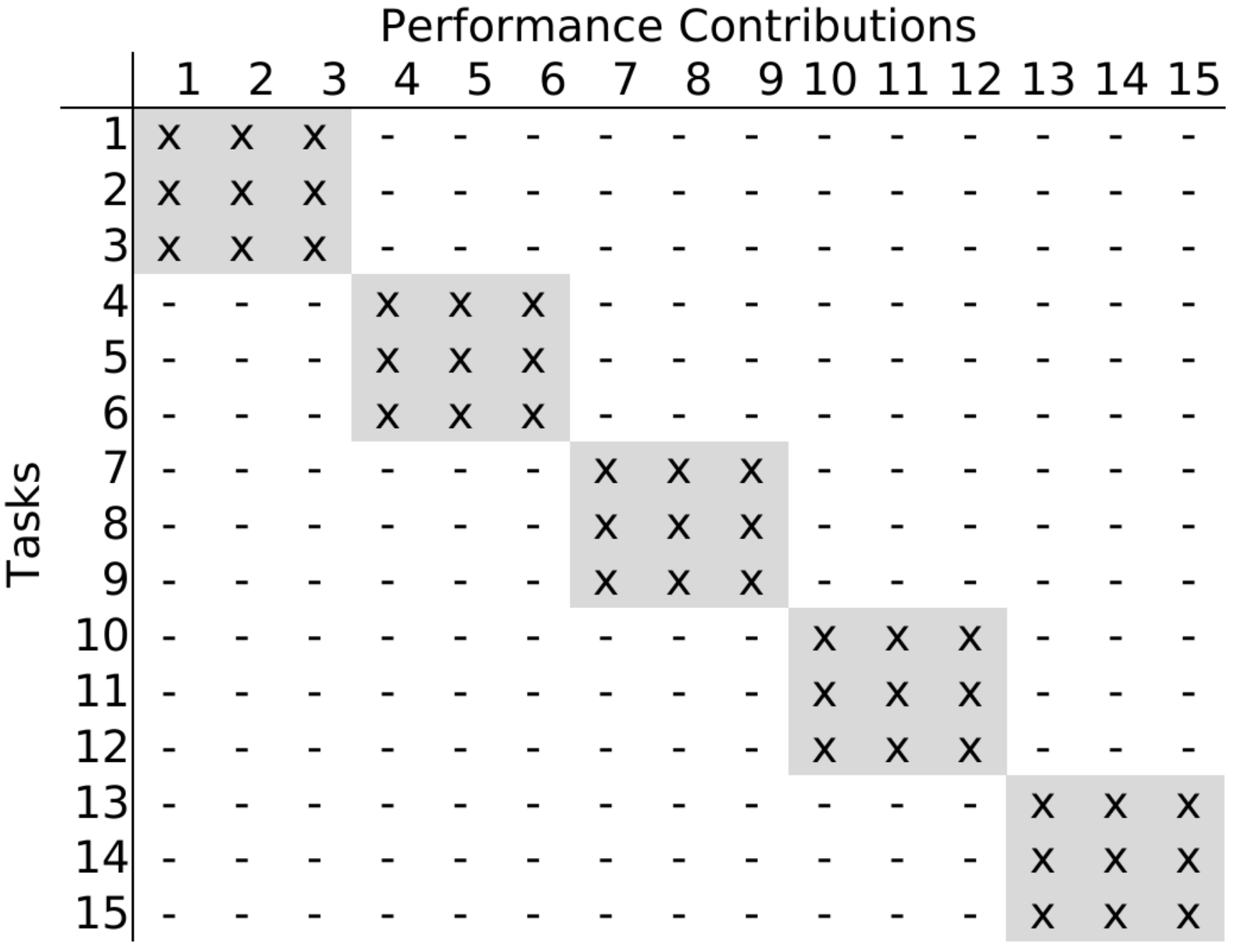}
        \caption{Modular}
    \end{subfigure}%
    \begin{subfigure}[t]{0.48\textwidth}
        \centering
         \label{fig:matrices-non-decomposable}
        \includegraphics[scale=0.305]{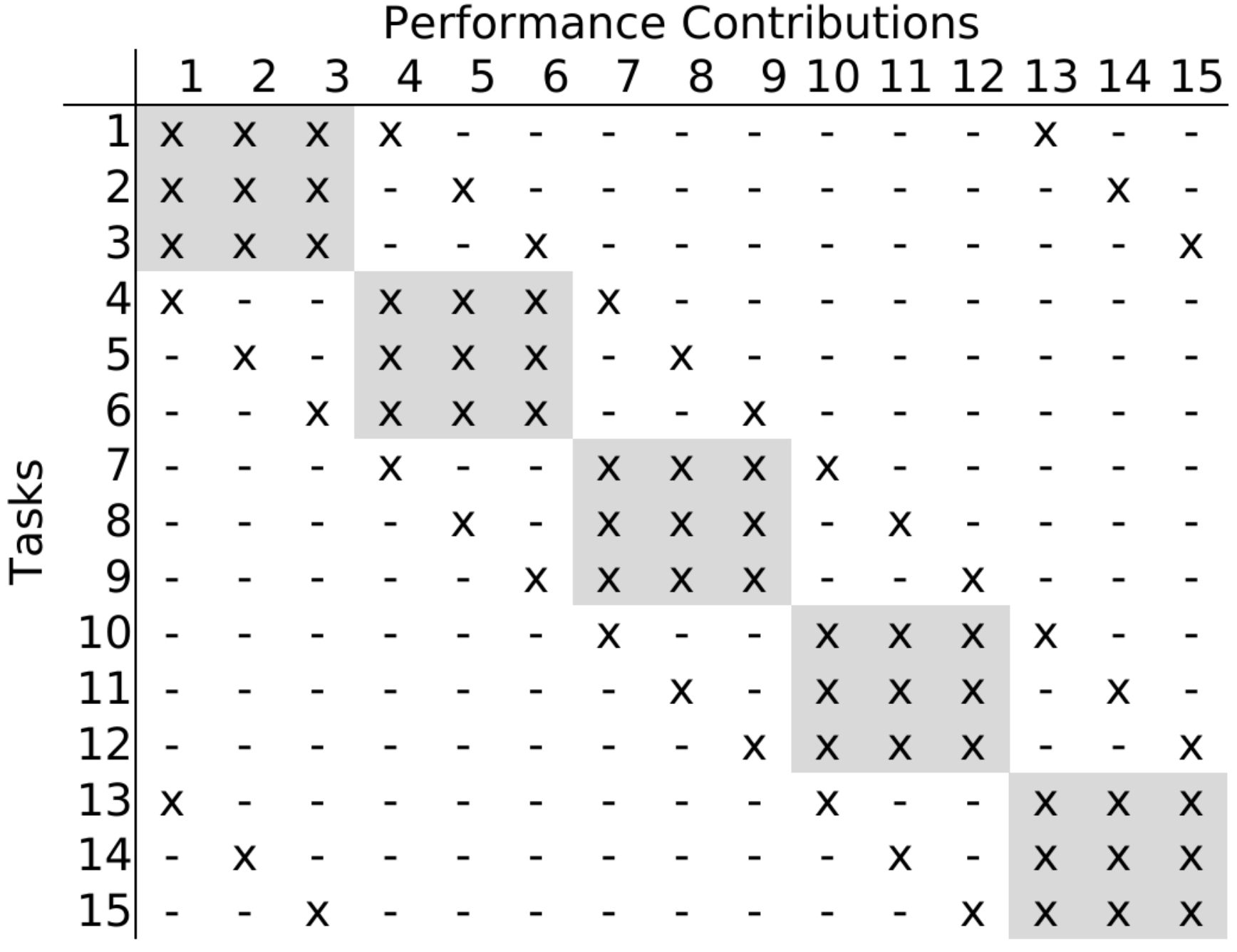}
        \caption{Non-modular}
    \end{subfigure}
    \caption{Interdependence patterns}
    \label{fig:matrices}
    \vspace{-4mm}
\end{figure*}
\begin{table}
\scriptsize
\caption{Parameters}
\label{tab:variables}
\setlength\extrarowheight{0pt} 
\begin{tabular*}{\textwidth}{@{\extracolsep{\fill}}*{4}{llll}}
\\ 
\toprule
Type                                    & Variables              & Notation                    & Values                         \\ \midrule
\multirow{5}{*}{Independent variables}  & Shock correlation        & $\rho$                         & \{-0.5, 0.5\}                       \\
                                        & Interdependence pattern               & Pattern &  \{modular, non-modular\}\\
                                        & Re-allocation weight            & $\gamma$                         & $\{0, 1\}$   \\
                                        & Incentive parameter  & $\lambda$                & $\{0.33, 1\}$ \\
                                        & Time  & $t$                & $\{0 : 1 : 200\}$ \\ \midrule
Dependent variable                      & Normalized task performance       & $\tilde{P}_t$           & $[0,1]$                     \\ \midrule
\multirow{7}{*}{Other parameters}       
                                        & Number of tasks    & $N$                         & 15                             \\
                                        & Number of agents   & $M$                         & 5                            \\                                  
                                        & Agents' cognitive capacities   & $C$                         & 7                            \\
                                        & Task re-allocation interval     & $\tau$                      & $ 20$       \\
                                        & Time until shock      &   --    &       $50$\\
                                        & Prediction error  & $\varepsilon$ & $N(0, 0.01)$ \\
                                        & Number of simulations  & $S$                         & 600                        \\
\bottomrule
\end{tabular*}%
\vspace{-4mm}
\end{table}

\subsection{Parameters and data analysis}
\label{sec:parameters}
%
This paper presents an analysis of the robustness of organizations  in the face of unexpected shocks. Specifically, this paper investigates the organizational performance of bottom-up designed organizations compared to their top-down counterparts. To achieve this, we compare the performance metrics of these organizations, placing particular emphasis on the following four variables:
\begin{enumerate}
    \item The analysis consider external shocks of different severity, i.e., the study investigates the impact of external shocks on organizational performance by introducing correlated $N\!K$ performance landscapes with correlations from $\rho \in \{-0.5, 0.5\}$ to simulate positive and negative shocks (see Eqs. \ref{eq:correlation} and \ref{eq:correlated-shock}).
    \item The paper explores the effect of different pieces of information on agents' autonomous task re-allocation decisions, as presented in Eqs. \ref{eq:delta-utility} to \ref{eq:delta-plus}. The analysis is simplified to the two extreme cases of $\gamma \in \{0,1\}$, which represent scenarios where agents prioritize mirroring and the impact of task re-allocation on utility, respectively.
    \item The paper examines the role of incentive mechanisms in regulating agent behavior, given that they are the organization's sole means of control. The analysis focuses on variations in the incentive parameter (see Eq. \ref{eq:utility}). Specifically, cases with altruistic mechanisms that place little ($\lambda = 0.33$) emphasis on an agent's residual performance, and individualistic mechanisms that only consider the performance generated within an agent's area of responsibility ($\lambda = 1$) are considered.
    \item The study analyzes landscapes with patterned interactions of varying complexity. We examine two specific cases: \textit{(a)} decomposable tasks with interactions patterns that enable a symmetric full modularization of tasks, with no interdependencies between subtasks, and \textit{(b)} non-decomposable tasks with reciprocal interdependencies that preclude full modularization due to the complexity of the tasks. The interaction patterns are illustrated in Fig. \ref{fig:matrices}. The shaded areas in Fig. \ref{fig:matrices} indicate the task allocation in the benchmark cases (with top-down task allocation). 
\end{enumerate}
%
In order to ensure that the results are comparable both within and across scenarios and simulation runs, the analysis adopts the average normalized performance as the primary measure of organizational performance. To compute the average normalized performance for a given scenario, performance achieved by the organization in simulation run $s \in \{1, \dots, S\}$ and period $t$ is normalized by the maximum achievable performance in that simulation run and at that point in time. Let us denote the maximum attainable performance in simulation run $s$ and period $t$ by $P^{\text{max}}_{st}$. The average normalized performance in period $t$ is then computed as follows:
\begin{equation}\label{eq:norm}
\Bar{P}_t=\frac{1}{S}\sum_{t=1}^{S}\frac{P(\mathbf{d}_t)}{P^{\text{max}}_{st}} ~.   
\end{equation}
The average normalized performances before and immediately after the shock event are presented in Tab. \ref{tab:shock-results}, while the performances after the organization had time to recover from the shock are presented in Tab. \ref{tab:results-recovery}. Moreover, Tab. \ref{tab:shock-results} reports the absolute difference between the performances attained at two different points in time, denoted by $\tau$ and $t$, as follows: $\Delta_{\tau:t} = \Bar{P}_{t} - \Bar{P}_{\tau}$, and Tab. \ref{tab:results-recovery} reports the corresponding relative difference, $\Delta^{\text{rel}}_{\tau:t} = (\Bar{P}_{t} - \Bar{P}_{\tau})/\Bar{P}_{\tau}$.

\section{Results}
\label{sec:results}
\subsection{Ability to absorb external shocks}
\label{sec:results-absorbing}
The results of scenarios implementing bottom-up task allocation, which account for parameter variations outlined in Sec. \ref{sec:parameters}, as well as the corresponding benchmark scenarios, are presented in Tab. \ref{tab:shock-results}. This table displays the normalized performances achieved immediately before and after an external shock in columns $\tilde{P}_{50}$ and $\tilde{P}_{51}$, respectively, as well as the effects of such shocks on organizational performance in column $\Delta_{50:51}$. Furthermore, the column labeled "Sign." offers insights into whether scenarios employing bottom-up task allocation differ significantly from benchmark scenarios in terms of the severity of shocks. 

The absolute performances attained prior to the occurrence of external shocks reflect the findings previously presented in research studies. In cases featuring both bottom-up and top-down task allocation, performances are observed to be higher for scenarios with modular task structures than those with non-modular task structures \cite{wall2018emergence}. Additionally, \cite{leitner2023collaborative} reports that the relative advantage of top-down task allocation is more evident when individualistic incentive mechanisms are effective within the organization. Finally, as proposed by \cite{leitner2023designing}, utility-based task re-allocation mechanisms outperform self-organized approaches that follow the mirroring hypothesis if individualistic incentives are in place.

Regarding the severity of shocks, the findings presented in Tab. \ref{tab:shock-results} indicate that moderate (positively correlated) shocks have less pronounced effects compared to more severe (negatively correlated) shocks. It is worth noting that the performances of top-down-designed organizations after a shock remain higher compared to those of bottom-up-designed organizations, which is consistent with previous research \cite{wall2018emergence}. This finding also corroborates Nissen's concept of static organizational fit \cite{nissen2014organization}, highlighting its strengths in steady conditions and weaknesses during major upheavals. It uncovers a trade-off between organizational efficiency and adaptability, mirroring similar dynamics observed in supply chains and food networks by de Arquer et al. \cite{de2022examining} and Karakoc and Konar \cite{karakoc2021complex}.  However, the outcomes also demonstrate that organizations utilizing traditional top-down task allocation experience considerably greater declines in performance in almost all cases than those employing bottom-up task allocation. Hence, the results suggest that organizations utilizing bottom-up task allocation demonstrate a greater ability to absorb the absolute effects of shocks, highlighting the importance of task allocation approaches in enhancing organizational resilience.

\begin{table}[!ht]
\scriptsize
\caption{Ability to absorb shocks}
\label{tab:shock-results}
\setlength\tabcolsep{0pt}
\setlength\extrarowheight{0pt} 
\begin{tabular*}{\textwidth}{@{\extracolsep{\fill}}*{11}{llllrrrrrrl}}
\\
\toprule
\multicolumn{4}{c}{Parameters} & \multicolumn{3}{c}{Bottom-up} & \multicolumn{3}{c}{Benchmark}  & \multirow{2}{*}{Sign.}\\
\cmidrule{1-4} 
\cmidrule{5-7}
\cmidrule{8-10}
Shock & Pattern & Task alloc. & Incentive & $\Bar{P}_{50}$ & $\Bar{P}_{51}$ & $\Delta_{50:51}$&   $\Bar{P}_{50}$ & $\Bar{P}_{51}$ & $\Delta_{50:51}$ \\
\midrule
\multirow{9}{*}{positive}    & \multirow{4}{*}{modular}     & \multirow{2}{*}{mirroring}    & altruistic  & 0.835	& 0.773	& -0.062	& 0.937	& 0.864	& \textcolor{red}{\textit{-0.073}}	& \oneS\\
                        &                                   &                               & individual      & 0.766	& 0.733	& -0.033	& 0.936	& 0.868	& \textcolor{red}{\textit{-0.069}}	& \twoS\\
                        \cmidrule{3-11}
                        &                                   & \multirow{2}{*}{utility}    & altruistic  & 0.856	& 0.805	& -0.051	& 0.937	& 0.864	& \textcolor{red}{\textit{-0.073}}	& \twoS\\
                        &                                   &                        &  individual      & 0.829	& 0.780	& -0.049	& 0.936	& 0.868	& \textcolor{red}{\textit{-0.069}}	& \twoS\\
                        \cmidrule{2-11}
                        & \multirow{4}{*}{non-modular}      & \multirow{2}{*}{mirroring}    & altruistic  &  0.723	& 0.681	& -0.042	& 0.742	& 0.711	& -0.032	& n.s.\\
                        &                                   &                       & individual     &  0.674	& 0.664	& -0.010	& 0.736	& 0.712	& \textcolor{red}{\textit{-0.024}}	& \oneS\\
                        \cmidrule{3-11}
                        &                                   & \multirow{2}{*}{utility}    & altruistic  &  0.720	& 0.687	& -0.033	& 0.742	& 0.711	& -0.032	& n.s.\\
                        &                                   &                       & individual     & 0.692	& 0.677	& -0.016	& 0.736	& 0.712	& {-0.024}	& n.s. \\
\midrule
\multirow{9}{*}{negative}& \multirow{4}{*}{modular}         & \multirow{2}{*}{mirroring}    & altruistic  &  0.837	& 0.676	& -0.161	& 0.939	& 0.760	& \textcolor{red}{\textit{-0.179}} & 	\oneS\\
                        &                                   &                       & individual     & 0.771	& 0.691	& -0.081	& 0.935	& 0.767	& \textcolor{red}{\textit{-0.168}} & 	\twoS\\
                        \cmidrule{3-11}
                        &                                   & \multirow{2}{*}{utility}    & altruistic & 0.853	& 0.700	& -0.154	& 0.939	& 0.760	& \textcolor{red}{\textit{-0.179}}	& \twoS \\
                        &                                   &                       & individual    & 0.832	& 0.694	& -0.138	& 0.935	& 0.767	& \textcolor{red}{\textit{-0.168}}	& \twoS \\
                        \cmidrule{2-11}
                        & \multirow{4}{*}{non-modular}      & \multirow{2}{*}{mirroring}    & altruistic & 0.728	& 0.644	& -0.084	& 0.737	& 0.665	& -0.072	& n.s.\\
                        &                                   &                       & individual    &  0.677	& 0.641	& -0.036	& 0.742	& 0.670	& \textcolor{red}{\textit{-0.072}}	& \twoS\\
                        \cmidrule{3-11}
                        &                                   & \multirow{2}{*}{utility}    & altruistic &  0.714	& 0.649	& -0.065	& 0.737	& 0.665	& \textcolor{red}{\textit{-0.072}}	& \oneS \\
                        &                                   &                       & individual    &  0.701	& 0.651	& -0.050	& 0.742	& 0.670	& \textcolor{red}{\textit{-0.072}}	& \twoS \\
\bottomrule
\end{tabular*}%
\\ 
Parameters: Shock: Correlation of shocks $\rho$; Pattern: Interdependence pattern; Task alloc.: Re-allocation weight $\gamma$, Incentive: Incentive parameter $\lambda$ (see Tab. \ref{tab:variables}).\\
Symbols: $\Bar{P}_t$: average normalized performance in period $t$ (Eq. \ref{eq:norm}); $\Delta_{\tau:t}$: absolute difference between normalized performances in periods $\tau$ and $t$.\\
Significance between cases with bottom-up allocation and benchmark scenarios is computed using a student's t-test: n.s.: $p>0.05$; \oneS: $p \leq 0.05$; \twoS: $p \leq 0.01$. \\
Red (italic) formatting indicates the scenarios in which shocks have significantly more severe effects.\\
\vspace{-4mm}
\end{table}
\begin{table}[!ht]
\scriptsize
\caption{Ability to recover from shocks}
\label{tab:results-recovery}
\setlength\tabcolsep{0pt}
\setlength\extrarowheight{0pt} 
\begin{tabular*}{\textwidth}{@{\extracolsep{\fill}}*{11}{lllcccrlcrl}}
\\
\toprule
\multicolumn{4}{c}{Parameters} & Before shock & \multicolumn{3}{c}{Recovery in $t=100$} & \multicolumn{3}{c}{Recovery in $t=200$}  \\
\cmidrule{1-4}
\cmidrule{5-5}
\cmidrule{6-8}
\cmidrule{9-11}
Shock & Pattern & Task alloc. & Incentive   & $\Bar{P}_{50}$ & $\Bar{P}_{100}$ & $\Delta^{\text{rel}}_{50:100}$& & $\Bar{P}_{200}$ & $\Delta^{\text{rel}}_{50:200}$ & \\
\midrule
\multicolumn{8}{l}{\textbf{Scenarios with bottom-up task allocation}} \\
\multirow{9}{*}{positive}    & \multirow{4}{*}{modular}       & \multirow{2}{*}{mirroring}  & altruistic  & 0.835		& 0.822	& \textcolor{red}{\textit{-1.60\%}}	& \twoS	& 0.826	& -1.06\%	 & n.s. \\
                        &                                   &                       & individual     & 0.766		& 0.770	&  0.55\%	& n.s.	& 0.769	&   0.40\%	 & n.s. \\
                        \cmidrule{3-11}
                        &                                   & \multirow{2}{*}{utility}    & altruistic  & 0.856		& 0.886	&  \textcolor{green}{\textbf{3.45\%}}	& \twoS	& 0.898	&   \textcolor{green}{\textbf{4.93\%}} &\twoS \\
                        &                                   &                       & individual     & 0.829		& 0.864	&  \textcolor{green}{\textbf{4.15\%}}	& \twoS	& 0.877	&   \textcolor{green}{\textbf{5.75\%}}	&\twoS \\
                        \cmidrule{2-11}
                        & \multirow{4}{*}{non-modular}      & \multirow{2}{*}{mirroring}    & altruistic  & 0.723		& 0.705	& \textcolor{red}{\textit{-2.48\%}}	& \twoS	& 0.724	& 0.10\%	 &	n.s. \\
                        &                                   &                       & individual     & 0.674		& 0.690	& \textcolor{green}{\textbf{2.38\%}}	& \twoS	& 0.684	& 1.50\%	 &	n.s. \\
                        \cmidrule{3-11}
                        &                                   & \multirow{2}{*}{utility}    & altruistic  & 0.720		& 0.711	& -1.28\%	&  n.s. 	& 0.714	& -0.82\%	 &	 n.s. \\
                        &                                   &                       & individual     & 0.692		& 0.697	& 0.67\%	&  n.s. 	& 0.702	& 1.41\%	 &	 n.s. \\
\midrule
\multirow{9}{*}{negative}& \multirow{4}{*}{modular}         & \multirow{2}{*}{mirroring}    & altruistic  & 0.837		& 0.820	& \textcolor{red}{\textit{-2.07\%}}	& \twoS	& 0.828	& -1.07\%	 &	 n.s. \\
                        &                                   &                       & individual     & 0.771	& 0.779	& 0.95\%	&  n.s. 	& 0.782	& \textcolor{green}{\textbf{1.41\%}}	 &	\oneS\\
                        \cmidrule{3-11}
                        &                                   & \multirow{2}{*}{utility}    & altruistic  & 0.853		& 0.876	& \textcolor{green}{\textbf{2.68\%}}	& \twoS	& 0.893	& \textcolor{green}{\textbf{4.63\%}}	 &	\twoS\\
                        &                                   &                       & individual     & 0.832		& 0.858	& \textcolor{green}{\textbf{3.13\%}}	& \twoS	& 0.879	& \textcolor{green}{\textbf{5.64\%}}	 &	\twoS\\
                        \cmidrule{2-11}
                        & \multirow{4}{*}{non-modular}      & \multirow{2}{*}{mirroring}    & altruistic  & 0.728		& 0.714	& \textcolor{red}{\textit{-1.98\%}}	& \oneS	& 0.716	& -1.63\%	 &	 n.s. \\
                        &                                   &                       & individual     & 0.677		& 0.674	& -0.39\%	&  n.s. 	& 0.679	& 0.33\%	 &	 n.s. \\
                        \cmidrule{3-11}
                        &                                   & \multirow{2}{*}{utility}    & altruistic  & 0.714		& 0.707	& -1.09\%	&  n.s. 	& 0.709	& -0.77\%	 &	 n.s. \\
                        &                                   &                       & individual     & 0.701		& 0.696	& -0.70\%	&  n.s. 	& 0.695	& -0.90\%	 &	 n.s. \\
\midrule
\multicolumn{8}{l}{\textbf{Benchmark scenarios with top-down task allocation}} \\
\multirow{4}{*}{positive}& \multirow{2}{*}{modular}         & \multirow{2}{*}{n.a.}    & altruistic & 0.937		& 0.946	& \textcolor{green}{\textbf{0.97\%}}	& \twoS	& 0.946 & 	\textcolor{green}{\textbf{0.97\%}}	& \twoS\\
                        &                                   &                       & individual    & 0.936		& 0.945	& \textcolor{green}{\textbf{0.94\%}}	& \twoS	& 0.945	& \textcolor{green}{\textbf{0.94\%}}	& \twoS\\
                        \cmidrule{2-11}
                        & \multirow{2}{*}{non-modular}      & \multirow{2}{*}{n.a.}    &  altruistic & 0.742	& 0.719	& \textcolor{red}{\textit{-3.08\%}}	& \twoS	& 0.719	& \textcolor{red}{\textit{-3.08\%}}	& \twoS\\
                        &                                   &                       &  individual    & 0.736		& 0.723	& \textcolor{red}{\textit{-1.78\%}}	& \oneS	& 0.723	& \textcolor{red}{\textit{-1.78\%}}	& \oneS\\
                        \midrule 
\multirow{4}{*}{negative}& \multirow{2}{*}{modular}         & \multirow{2}{*}{n.a.}    &  altruistic &0.939		& 0.931	& \textcolor{red}{\textit{-0.83\%}}	& \twoS & 	0.931	& \textcolor{red}{\textit{-0.83\%}}	& \twoS\\
                        &                                   &                       & individual     &0.935		& 0.931	& -0.45\%	& n.s.	& 0.931 & 	-0.45\%	& n.s.\\
                        \cmidrule{2-11}
                        & \multirow{2}{*}{non-modular}      & \multirow{2}{*}{n.a.}    & altruistic &0.737	& 0.716	& \textcolor{red}{\textit{-2.86\%}}	& \twoS	& 0.716	& \textcolor{red}{\textit{-2.86\%}}	& \twoS\\
                        &                                   &                       & individual     & 0.742		& 0.705	& \textcolor{red}{\textit{-5.09\%}}	& \twoS	& 0.705	& \textcolor{red}{\textit{-5.09\%}}	& \twoS\\
                        \bottomrule
\end{tabular*}%
\\ 
Parameters: Shock: Correlation of shocks $\rho$; Pattern: Interdependence pattern; Task alloc.: Re-allocation weight $\gamma$, Incentive: Incentive parameter $\lambda$; n.a. indicates that the parameter is not applicable in the scenario (see Tab. \ref{tab:variables})\\
Symbols: $\Bar{P}_{50}$: average normalized performance in period $t$ (Eq. \ref{eq:norm}); $\Delta^{\text{rel}}_{\tau:t}$: relative difference between normalized performances in periods $\tau$ and $t$.\\
Significance is computed using a paired t-test: n.s.: $p>0.05$; \oneS: $p \leq 0.05$; \twoS: $p \leq 0.01$. \\
Bold (green) and italic (red) font indicates that the performance is significantly higher and lower, respectively, compared to the pre-shock performance.
\vspace{-5mm}
\end{table}
\subsection{Ability to recover from external shocks}
\label{sec:results-recovering}
Section \ref{sec:results-absorbing} focuses on the immediate effects of external shocks on an organization's performance. In contrast, this section examines the organization's ability to recover from such disturbances in its environment. Table \ref{tab:results-recovery} presents the attained performance before ($\tilde{P}_{50}$) and after $100$ ($\tilde{P}_{100}$) and $200$ periods ($\tilde{P}_{200}$). The resulting recovery metrics are presented in terms of the relative performance levels after $100$ and $200$ periods, represented by the columns $\Delta^{\text{rel}}_{50:100}$ and $\Delta^{\text{rel}}_{50:200}$, respectively, compared to the performance before the shock. Additionally, bold (green) and italic (red) formatting is utilized to indicate significantly higher and lower attained performance levels, respectively, relative to the pre-shock performance.
\subsubsection{Scenarios with bottom-up task allocation}
In scenarios featuring bottom-up task allocation, the mode of task re-allocation emerges as a critical design parameter in fostering an organization's resilience to shocks. Specifically, when agents make task re-allocation decisions based on their personal utility instead of aligning them with the logic of the mirroring hypothesis, an organization's performance after a shock is significantly improved compared to the performance \textit{before} the shock. This finding holds true across analyzed incentive parameters and for both moderate (positively correlated) and more severe (negatively correlated) shocks. The result is consistent in the short term (50 periods after a shock) and becomes more pronounced over the long term (100 periods after a shock). Conversely, if agents align their re-allocation decisions with the mirroring hypothesis, they can still recover from shocks in most cases and attain pre-shock performance levels.
When organizations encounter a non-modular task structure, they exhibit the ability to recover from environmental disruptions in all cases over the long term, with organizations achieving pre-shock performance after 200 periods. However, results suggest that achieving recovery in the short term (50 periods after a shock) takes longer when task re-allocation is based on mirroring and incentives are altruistic. In such instances, the performance level attained 50 periods after the shock had not yet reached pre-shock levels.
\subsubsection{Scenarios with top-down task allocation}
In scenarios featuring top-down task allocation, an organization's ability to recover from environmental disruptions is significantly influenced by the interplay between the task structure and the severity of the shock. Specifically, moderate (positively correlated) shocks and tasks with a modular interdependence structure allow organizations to improve their performance levels relative to the pre-shock performance even in the short term. However, when faced with a more severe (negatively correlated) shock and a modular task structure, an organization's recovery is contingent on the incentive mechanism in place, with only individualistic incentives enabling the organization to attain pre-shock performance. Moreover, when the task structure features a non-modular interdependence, organizations cannot fully recover from shocks, meaning that neither in the short nor long term can they achieve pre-shock performance levels, with this result holding true for both moderate (positively correlated) and more severe (negatively correlated) shocks.

\subsubsection{Comparison between scenarios}
The findings, as presented in Tab. \ref{tab:results-recovery}, reveal that top-down designed organizations hold an advantage in the recovery process only under certain conditions. Specifically, such organizations are better positioned to recover if the interdependence structure underlying the task at hand is modular, if task re-allocation follows the mirroring hypothesis, and if the organization experiences a moderate shock.
However, in all other cases, the ability of top-down designed organizations to recover from shocks and retain pre-shock performance is lower when compared to their bottom-up designed counterparts. For instance, when the task structure is modular and the shock is severe, top-down designed organizations exhibit a performance that is similar to, or slightly below, the pre-shock performance. In contrast, bottom-up designed organizations are able to achieve a performance that exceeds pre-shock performance or is, at the very least, equivalent to pre-shock performance. Similarly, in the case of non-modular task structures, bottom-up designed organizations are able to restore their performance to pre-shock levels, whereas top-down designed organizations exhibit a performance that is significantly lower than pre-shock levels.

These findings reaffirm the vital role of social connections (and collaboration) in organizational resilience, as highlighted in prior studies like Su and Junge \cite{su2023unlocking}, with a special focus on the benefits of internal collaborative dynamics. It demonstrates that inter-departmental collaboration within organizations facilitates both recovery from disruptions and performance enhancement, particularly when incorporating dynamic task allocation. Additionally, the study draws parallels with firms' strategies during the COVID-19 pandemic---i.e., boosted employee communication and collaboration with remote-working technologies, resulting in performance increases---, as, amongst others, investigated by Guan et al. \cite{guan2023organizational}.
\section{Conclusions}
\label{sec:conclusion}

This paper extends previous research on organizational adaptation and design \cite{leitner2023designing} by including a more fine-grained model of motives during task allocation and a model of exogenously controllable environmental disruptions. The results of this study provide valuable insights into the resilience of bottom-up designed organizations, specifically the impact of micro-level task allocation decisions on the macro-level resilience of an organization. The findings indicate that a bottom-up designed structure enhances an organization's ability to absorb adverse effects triggered by environmental disruptions. Furthermore, the results suggest that while top-down designed organizations may have advantages in certain contexts, the flexibility and decentralized nature of bottom-up designed organizations may enable them to recover from disruptions in a wider range of scenarios. 

The results presented in this paper focus exclusively on the resilience of a single organization. However, disruptions in the environment can have a ripple effect throughout the entire supply chain, affecting interdependent organizations. Therefore, future research could explore the resilience of bottom-up designed organizations within a network of interdependent organizations. Furthermore, the model presented in this paper assumes that disruptions in the environment only affect the performance contributions, while the interdependence patterns remain unchanged. However, future research could extend the effects of shocks to changes in the interdependence patterns as well.

%
%
%
\bibliography{bib}

\begin{thebibliography}{10}
\providecommand{\url}[1]{\texttt{#1}}
\providecommand{\urlprefix}{URL }

\bibitem{de2022examining}
de~Arquer, M., Ponte, B., Pino, R.: Examining the balance between efficiency
  and resilience in closed-loop supply chains. Central European Journal of
  Operations Research  30,  1307--1336 (2022)

\bibitem{branicki2019resilience}
Branicki, L., Steyer, V., Sullivan-Taylor, B.: Why resilience managers aren’t
  resilient, and what human resource management can do about it. The
  International Journal of Human Resource Management  30(8),  1261--1286 (2019)

\bibitem{christopher2004supplychain}
Christopher, M., Peck, H.: Building the resilient supply chain. The
  International Journal of Logistics Management  15,  1--14 (2004)

\bibitem{colfer2016mirroring}
Colfer, L.J., Baldwin, C.Y.: The mirroring hypothesis: theory, evidence, and
  exceptions. Industrial and Corporate Change  25(5),  709--738 (2016)

\bibitem{demirtas2014generating}
Demirtas, H.: Generating bivariate uniform data with a full range of
  correlations and connections to bivariate binary data. Communications in
  Statistics-Theory and Methods  43(17),  3574--3579 (2014)

\bibitem{fiksel2015}
Fiksel, J., Polyviou, M., K.L., C., Pettit, T.J.: From risk to resilience:
  Learning to deal with disruption. MIT Sloan Management Review  56,  79--86
  (2015)

\bibitem{guan2023organizational}
Guan, F., Tienan, W., Tang, L.: Organizational resilience under covid-19: the
  role of digital technology in r\&d investment and performance. Industrial
  Management \& Data Systems  123,  41--63 (2023)

\bibitem{hepfer2022heterogeneity}
Hepfer, M., Lawrence, T.B.: The heterogeneity of organizational resilience:
  Exploring functional, operational and strategic resilience. Organization
  Theory  3 (2022)

\bibitem{hollnagel2011resilienceengineering}
Hollnagel, E.: The scope of resilience engineering. In: Hollnagel, E.,
  Pari\'{e}s, J., Woods, D.D., Wreathal, J. (eds.) Resilience engineering in
  practice: A guidebook, pp. XXIXE -- XXXXIX. CRC Presss, Taylor \& Francis
  Group (2011)

\bibitem{karakoc2021complex}
Karakoc, D.B., Konar, M.: A complex network framework for the efficiency and
  resilience trade-off in global food trade. Environmental Research Letters  16
  (2021), paper 105003

\bibitem{leitner2023collaborative}
Leitner, S.: Collaborative search and autonomous task allocation in
  organizations of learning agents. In: Squazzoni, F. (ed.) Advances in Social
  Simulation, ESSA2022, pp. 345--357. Springer Proceedings in Complexity,
  Springer Nature Switzerland AG (2023)

\bibitem{leitner2023designing}
Leitner, S.: Designing organizations for bottom-up task allocation: The role of
  incentives (2023), arXiv.2301.00410

\bibitem{leitner2015simulation}
Leitner, S., Wall, F.: Simulation-based research in management accounting and
  control: an illustrative overview. Journal of Management Control  26,
  105--129 (2015)

\bibitem{linnenluecke2017resilience}
Linnenluecke, M.K.: Resilience in business and management research: A review of
  influential publications and a research agenda. International Journal of
  Management Reviews  19(1),  4--30 (2017)

\bibitem{mcewen2018measure}
McEwen, K., Boyd, C.M.: A measure of team resilience: Developing the resilience
  at work team scale. Journal of Occupational and Environmental Medicine
  60(3),  258--272 (2018)

\bibitem{meyer1982adapting}
Meyer, A.D.: Adapting to environmental jolts. Administrative Science Quarterly
  pp. 515--537 (1982)

\bibitem{nissen2014organization}
Nissen, M.E.: Organization design for dynamic fit: A review and projection.
  Journal of Organization Design  3,  30--42 (2014)

\bibitem{ponomarov2009understanding}
Ponomarov, S.Y., Holcomb, M.C.: Understanding the concept of supply chain
  resilience. The International Journal of Logistics Management  20,  124--143
  (2009)

\bibitem{prayag2020psychological}
Prayag, G., Spector, S., Orchiston, C., Chowdhury, M.: Psychological
  resilience, organizational resilience and life satisfaction in tourism firms:
  Insights from the canterbury earthquakes. Current Issues in Tourism  23(10),
  1216--1233 (2020)

\bibitem{raetze2021resilience}
Raetze, S., Duchek, S., Maynard, M.T., Kirkman, B.L.: Resilience in
  organizations: An integrative multilevel review and editorial introduction.
  Group \& Organization Management  46(4),  607--656 (2021)

\bibitem{staw1981threat}
Staw, B.M., Sandelands, L.E., Dutton, J.E.: Threat rigidity effects in
  organizational behavior: A multilevel analysis. Administrative Science
  Quarterly pp. 501--524 (1981)

\bibitem{su2023unlocking}
Su, W., Junge, S.: Unlocking the recipe for organizational resilience: A review
  and future research directions. European Management Journal  (2023), in press

\bibitem{wall2018emergence}
Wall, F.: Emergence of task formation in organizations: Balancing units'
  competence and capacity. Journal of Artificial Societies and Social
  Simulation  21(2),  1--26 (2018)

\bibitem{wall2021agent}
Wall, F., Leitner, S.: Agent-based computational economics in management
  accounting research: Opportunities and difficulties. Journal of Management
  Accounting Research  33(3),  189--212 (2021)

\bibitem{youssef2005}
Youssef, C.M., Luthans, F.: Resiliency development in organiztions, leaders and
  employees: Multi-level theory building for sustained performance. In:
  Gardner, W.L., Aviolo, B.J., Walumbwa, F.O. (eds.) Authentic Leadership
  Theory and Practice: Origins, Effects, and Development, pp. 303--343.
  Elsevier (2005)

\end{thebibliography}
\bibliographystyle{splncs03}

\end{document}